\documentclass[12pt]{iopart}
\usepackage{iopams} 
\usepackage[latin1]{inputenc}
\usepackage{graphicx}
\usepackage{url}

\begin{document}

\title{Dispersion properties of plasmonic sub-wavelength elliptical wires wrapped with graphene}


\author{Mauro Cuevas$^1$$^2$} 
\address{$^1$Consejo Nacional de Investigaciones Cient\'ificas y T\'ecnicas (CONICET)}
\address{$^2$ Universidad Austral, Facultad de Ingenier\'ia, Pilar Mariano Acosta 1611 B1629WWA-Pilar-Buenos Aires
Argentina}
\ead{mcuevas@austral.edu.ar}
\author{Ricardo A. Depine$^3$} 
\address{$^3$Grupo de Electromagnetismo Aplicado, Departamento de F\'isica, FCEN, Universidad de Buenos Aires and IFIBA, Ciudad Universitaria, Pabell\'on I, C1428EHA, Buenos Aires, Argentina}

\begin{abstract}
One fundamental motivation to know the dispersive, or frequency dependent characteristics  of localized surface plasmos (LSPs) supported by elliptical shaped particles wrapped with graphene sheet, as well as their  scattering  characteristics when these elliptical LSPs are excited,  is  related with the design of plasmonic structures capable to manipulate light at sub-wavelength scale. 
The anisotropy imposed by the  ellipse eccentricity can be used as a geometrical tool for controlling plasmonic resonances. 
Unlike metallic case, where the multipolar eigenmodes are independent of each others, we find that the induced  current on graphene boundary couples multipolar eigenmodes with the same parity. In the long wavelength limit, a  recursive relation equation for LSPs in term of the  ellipse eccentricity parameter is derived  and explicit solutions at lowest order are presented. 
In this approximation, we obtain analytical expressions for both the anisotropic polarizability tensor elements and the scattered power when LSPs are excited by plane wave incidence.        
\end{abstract}

\section{Introduction} \label{introduccion}

Light scattering properties on metallic wire particles with  non-circular cross section  have been extensively studied  \cite{Bohren,Maier}. The  two-dimensional geometrical anisotropy leads to a splitting in two or more plasmon branches corresponding to the lowest energy states, instead of one as occur in the circular case, which are evidenced on the  angular optical response of the wire \cite{Klimov}. This fact together with other  outstanding properties 
have found applications in some optical topics requiring  field hot spots, such as surface enhanced Raman and optical nano-antennas development  \cite{martin,Kneipp,rogobete1,lalanne}.

Sub-wavelength structures coated with a graphene sheet provide a suitable alternative to metallic elements because they exhibit relative  low loss and highly tunable, via electrostatic gating or chemical doping \cite{Xia}, surface plasmons in the frequency region from microwaves  to  infrared \cite{jablan}. The 
small plasmon wavelength, which usually reaches values smaller than one tenth of the wavelength of the photon of the same frequency,  
deals with the possibility to build smaller plasmonic constituent elements, 
a feature positioning the graphene as a promising platform to the development of controllable plasmon devices, in particular of a new generation of sensors and modulators   from microwaves to the mid-infrared regimes \cite{azzaroni,karanikolas,koppens,ShivaSR} constituting an intersection between optics and  electronic.

Graphene layers structures has attracted wide attention in many applications due to a strong adsorption capacity for contaminants \cite{cita1,cita2}, electrical properties \cite{cita4,cita_small,cita_material_views} and improved light harvesting \cite{miao,du,ni}.

In the framework of electromagnetic scattering by sub-wavelength graphene particles, an extensive wealth of theoretical analysis have been developed, allowing possible a wide range of applications based on the interaction between graphene and electromagnetic radiation via LSP mechanisms,   including sensing \cite{velichko,farmani}, superscattering \cite{shiva,gingins}, low energy  spasers \cite{Berman,Faez,leila,CKZ}, PT-symmetric structures tailoring lasing modes \cite{Zhang,CKZ}, 
and micro and nano antennas \cite{Jornet,CorreasSerrano,CuevasAntenas,Ullah}.

This work deals with the plasmon properties of an elliptical dielectric wire wrapped with graphene. The  LSP branches corresponding to lower eigenmodes are obtained by solving the homogeneous problem, \textit{i.e.}, the scattering problem without external excitation.  Unlike the metallic case, where LSP eigenmodes are uncoupled between them, the current density on graphene coating couples all multipolar LSPs with the same parity given rise to an infinite set of coupled equations for the field amplitude. Moreover, the tangential direction of charge  oscillation on graphene covered leads to a counterintuitive splitting: the low dipolar frequency mode corresponds to  polarizability oscillations along minor ellipse axis  while the high dipolar frequency mode corresponds to polarizability oscillations along the major ellipse axis.  In this framework, analytical expressions of the anysotropic polarizability is found. We also provided analytical expressions for the scattered power when the graphene elliptical wire is excited by a plane wave. All scattering curves are compared with those obtained by applying a  rigorous formalism based in the Green surface integral method \cite{valencia,cuevas6} valid for arbitrary shaped cross sections. 

This paper is organized as follows. Firstly, in Section 2 we develop an analytical method based on the separation of variables  in elliptical coordinates and obtain an approximated solution for the electromagnetic field scattered by an elliptical wire cover with graphene. This approach allows us to express each of the  amplitudes in the multipolar expansion of the fields as a series of power of the ellipse eccentricity. 
In Section 3 we present the results related with the dispersive characteristics  and the  scattering calculations for a dielectric wire wrapped with graphene. Finally, concluding remarks are provided in Section 4. The Gaussian system of units is used and an $\mbox{exp}(-i\omega t)$ time-dependence is implicit throughout the paper, with $ \omega$ as the angular frequency, $t$ as the time, and $i =  \sqrt{-1} $. The symbols Re and Im are respectively used for denoting the real and imaginary parts of a complex quantity.

\section{Theory}
\subsection{Scattered field equations}
We consider an elliptical dielectric  cylinder wrapped with graphene sheet. The elliptical profile has the  major semi-axis $a$  along $x$ axis and the minor semi-axis $b$ along the $y$ axis.  We use  elliptical coordinates, which are related with the Cartesian coordinates by following relations
\begin{equation}\label{coordenadas}
\begin{array}{ll}
x = l \cosh(\rho) \cos(\phi),\\
y= l \sinh{}(\rho) \sin(\phi),\\
z=z
\end{array}
\end{equation}
 $\rho$ and $\phi$ are the radial and angular elliptical coordinates. The unit vectors  $\hat{\rho}$ and $\hat{\phi}$ are normal and tangent along elliptical shape, respectively (see Figure 1).  
In this way, the boundary curve of the ellipse  is given by $\rho=\rho_0$,  $\tanh(\rho_0)=b/a$, with the major semi-axis $a=l \cosh(\rho_0)$ along the $x$ axis and the minor semi-axis $b=l \sinh(\rho_0)$ along the $y$ axis. The scale parameter $l$ is related with $a$ and $b$ by the following  $l=\sqrt{a^2-b^2}$. 

\begin{figure}
\centering
\resizebox{0.7\textwidth}{!}
{\includegraphics{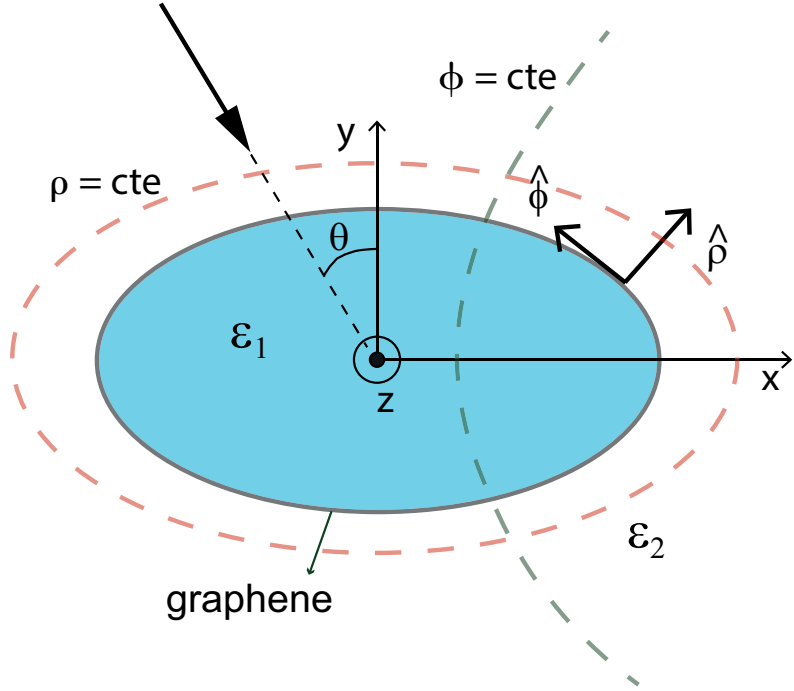}}
\caption{\label{fig:epsart} Schematic illustration of the elliptical wire wrapped with graphene. The wire ($\varepsilon_1$ and surface conductivity $\sigma$) is embedded in a
transparent medium with permittivity $\varepsilon_2$. Both media are non magnetic, $\mu_1=\mu_2=1$. 
}\label{sistema}
\end{figure}

The magnetic  field inside  the elliptic  cylinder $\rho<\rho_0$ can be expressed  as follows:
%
\begin{equation}\label{campo_dentro}
H^{(1)}(\rho,\phi) =
\sum_{m=1}^{+\infty} a_m \sinh(m \rho) \sin(m \phi) + c_m \cosh(m \rho) \cos(m \phi), 
\end{equation}
and the field outside the ellipse $\rho>\rho_0$  is expressed as,
\begin{equation}\label{campo_fuera}
H^{(2)}(\rho,\phi) =  \sum_{m=1}^{+\infty} b_m e^{-m \rho}  \sin(m \phi)+ d_m e^{-m \rho}  \cos(m \phi)  +H^{(inc)}.
\end{equation}
where the last term corresponds to the incident magnetic field. 
Firstly, we suppose a plane wave incidence with the electric field along the $x$ axis and  amplitude $E_0$. In this case,  the incident magnetic field  in the long wavelength limit is written as 
\begin{equation}\label{campo_magnetico_inc}
H^{(inc)}(\rho,\phi) =
-i k_0 E_0 \varepsilon_2 l \sinh(\rho) \sin(\phi). 
\end{equation}
The symmetry imposed by the incident field direction imposes the amplitudes $c_m$ and $d_m$ in the field expressions (\ref{campo_dentro}) and (\ref{campo_fuera}) to be zero. 
By using the Ampere-Maxwell equation, we can relate the components of the electric field and the $z$-component of the magnetic field as follows,
\begin{equation}\label{campo_electrico}
\mathbf{E}^{(j)} = -\frac{1}{ i k_0 \varepsilon_j} \mathbf{\nabla}_t \times \hat{z} H^{(j)}
\end{equation}
 $j=1,\,2$,   
\begin{equation}\label{campo_electrico1}
\mathbf{\nabla}_t = \frac{1}{ f(\rho_0,\phi)} \left[ \hat{\rho} \frac{\partial}{\partial \rho}+\hat{\phi} \frac{\partial}{\partial \phi}\right]
\end{equation}
is the transverse part of the $\mathbf{\nabla}$ operator and $f(\rho_0,\phi)=  l  \sqrt{\cosh^2(\rho)-\cos^2(\phi)}$.  
The boundary condition along the contour of the ellipse  $\rho=\rho_0$ is written as \cite{RCD}, 
\begin{equation}\label{CC}
\begin{array}{ll}
\frac{1}{\varepsilon_1} \frac{\partial H^{(1)}}{\partial  \rho}|_{\rho_0}=\frac{1}{\varepsilon_2} \frac{\partial H^{(2)}}{\partial  \rho}|_{\rho_0}\\
\left[H^{(2)}+H^{(inc)}-H^{(1)}\right]|_{\rho_0}=\frac{4\pi \sigma }{c 
} 
[E^{(2)}_\phi + E^{(inc)}_\phi] |_{\rho_0}.  
\end{array}
\end{equation}
Using field expressions (\ref{campo_dentro}) and (\ref{campo_fuera}) with $c_m=d_m=0$  into the boundary condition (\ref{CC}) and projecting into the Fourier basis, two system of equations 
for amplitudes $a_m$ and $b_m$ are obtained. Then, 
we can eliminate amplitudes $a_m$ to obtain one set of equations for amplitudes $B_m=b_m e^{-m \rho_0}$  (see appendix  \ref{desarrollo1}), 
\begin{eqnarray}\label{nohomogeneo}
 B_k t_k +  \sum_{m\not=k}^{+\infty}  M_{km} B_m   =Q_k, 
\end{eqnarray}
with
\begin{eqnarray}\label{M}
 M_{km} = \frac{4 \pi \sigma}{c k_0 \varepsilon_2}i  m s_{km}, 
\end{eqnarray}
\begin{eqnarray}\label{t}
 t_{k} = 1+\frac{\varepsilon_1}{\varepsilon_2} \tanh(k \rho_0)+M_{kk},
\end{eqnarray}
%
%
%
%
\begin{eqnarray}\label{Qk}
 Q_{k} = i k_0 E_0 (\varepsilon_2-\varepsilon_1) l \sinh(\rho_0) \delta_{k 1}+ \frac{4\pi \sigma }{c} E_0  \coth(\rho_0) s_{k1}.
\end{eqnarray}

\subsection{Lowest order eccentricity}

Before studying the scattering problem given by Eq. (\ref{nohomogeneo}), we first study the homogeneous problem (or eigenmodes problem), \textit{i.e.}, the scattering problem without an external source ($E_0=0$ in Eq. (\ref{nohomogeneo})). To do this, we consider the homogeneous part in Eq. (\ref{nohomogeneo}) which can be rewritten as
\begin{eqnarray}\label{homogeneo}
 B_k t_k +  \sum_{m\not=k}^{+\infty}  M_{km} B_m   =0, 
\end{eqnarray}
%
%
%
%
%
%
%
Since in the limit of null eccentricity $\tanh(k\rho_0)=1$ for $\rho=\infty$, the first term in Eq. (\ref{homogeneo}) reduces  to the dispersion relation of a cylinder with circular cross section for the $k$ multipolar order, while the second term reduce to zero. For small values of eccentricity, $e=l/b=\sqrt{(a/b)^2-1}<<1$, we can expand Eq. (\ref{homogeneo}) in powers of $e$ as follows. Taking into account that the matrix element is of order $O(2)$  for  $k\not=m$ (see appendix \ref{orden}), $s_{km}\approx e^2$, from 
Eq. (\ref{M}) we deduce that $M_{1m}$ is at last of order $O(2)$ and thus we can expand the homogeneous equation (\ref{homogeneo}) in powers of $e$. The form that Eq. (\ref{homogeneo}) takes for $k=1$ is,
\begin{eqnarray}\label{homogeneo2}
 B_1 t_1 +  \sum_{m\not=1}^{+\infty}  M_{1m} B_m   =0, 
\end{eqnarray}
and the form that Eq. (\ref{homogeneo}) takes for $m\not=1$ is,
\begin{eqnarray}\label{homogeneo3}
 B_m  =  -\frac{M_{m1}}{t_m}B_1- \sum_{n\not=1,m}^{+\infty}  \frac{M_{mn}}{t_m} B_n. 
\end{eqnarray}
Replacing Eq. (\ref{homogeneo3}) into Eq. (\ref{homogeneo2}), we obtain
\begin{eqnarray}\label{homogeneo4}
B_1 t_1-  \sum_{m\not=1}^{+\infty} M_{1m} \frac{M_{m1}}{t_m}B_1- \sum_{m\not=1}^{+\infty}   \sum_{n\not=1,m}^{+\infty} M_{1m} \frac{M_{mn}}{t_m} B_n   =0.
\end{eqnarray}

Following the same steps as described above, we solve Eq. (\ref{homogeneo}) by iteration,
\begin{eqnarray}\label{homogeneo4b}
 t_1  -\sum_{m\not=1}^{+\infty}  M_{1m}\frac{M_{m1}}{t_m} -\sum_{m\not=1}^{+\infty} 
 \sum_{n\not=1,m}^{+\infty} M_{1m}\frac{M_{mn}}{t_m} \frac{M_{n1}}{t_n}+...=0.
\end{eqnarray}
This equation is the dispersion relation of elliptical  LSPs on graphene. To find the LSP characteristics, we consider the lowest order, $e^4$, in the dispersion relation (\ref{homogeneo4b}),
\begin{eqnarray}\label{homogeneo5}
 t_1  - M_{13}\frac{M_{31}}{t_3}=0,
\end{eqnarray}
which can be made explicit using Eqs. (\ref{M}) and (\ref{t}) as follows
\begin{eqnarray}\label{homogeneo5b}
 \left(1+\frac{\varepsilon_1}{\varepsilon_2}\tanh(\rho_0)+\frac{4\pi\sigma}{c k_0 \varepsilon_2} i s_{11}\right) \nonumber \\ 
 \times \left(1+\frac{\varepsilon_1}{\varepsilon_2}\tanh(3\rho_0)+ 
 3 \frac{4\pi\sigma}{c k_0 \varepsilon_2} i s_{33}\right) + 3 \left( \frac{4\pi \sigma}{c k_0 \varepsilon_2}\right)^2 (s_{13})^2=0,
\end{eqnarray}
where we have used $s_{13}=s_{31}$ and the fact that $s_{12}=s_{21}=0$. 
The expressions for $s_{1m}$ and $s_{33}$ until order $e^2$ can be obtained from Eq. (\ref{s_1k}) (see appendix \ref{orden}), 
\begin{eqnarray}\label{s_1k2}
s_{11}=\frac{1}{b} \left[ 1-\frac{3}{8} e^2\right],\nonumber\\
s_{13}=\frac{1}{b} \frac{1}{8} e^2,\nonumber\\
s_{33}=\frac{1}{b} \left[ 1-\frac{1}{4} e^2\right].
\end{eqnarray}
We note that for null eccentricity, the matrix element $s_{13}=0$ and $\tanh(\rho_0)=\tanh(3\rho_0)=1$, as a  consequence the dispersion relation (\ref{homogeneo5b}) reduces to a product between two factors, both corresponding to a circular cylinder. One of these factors corresponds to the dispersion relation of the dipolar order ($m=1$) and the other corresponds to the hexapolar order ($m=3$). For small values of eccentricity, the matrix element   $s_{13}\not=0$ and consequently the last term in Eq. (\ref{homogeneo5b}) is not null, with which Eq. (\ref{homogeneo5b})  represents 
the 
elliptical LSP dispersion relation. 
Note that 
the coupling mechanism between the dipolar and the hexapolar orders 
is evidenced by the presence of the last term in the dispersion equation (\ref{homogeneo5b}). 

It is worth nothing that  Eq. (\ref{homogeneo5b}) is the lowest order of the elliptical LSP dispersion relation. 
This is true because the truncation of Eq. (\ref{homogeneo4}) at order $O(4)$ in eccentricity admit only modes $m \leq 4$. In addition, for higher orders,  $O(2N)$ for example, Eq. (\ref{homogeneo4}) couples $m=1, \,3,\, 5,...,2N-1$ multipolar orders. 

At the same order in eccentricity as we have written Eq. (\ref{homogeneo5}), from Eq. (\ref{nohomogeneo}) we can obtain the dipolar coefficient $b_1$ for the scattering of a plane wave (non homogeneous problem) polarized along $x$ axis, 
\begin{equation}\label{b1}
b_1 =E_0 e^{\rho_0}  \widetilde{b_1}, 
\end{equation}
where
\begin{eqnarray}
\widetilde{b_1} =
\frac{A_x}{t_1 t_3 - M_{13} M_{31}},
\end{eqnarray}
and
\begin{eqnarray}
A_x=
\left(ik_0 b [\varepsilon_2-\varepsilon_1]+\frac{4 \pi \sigma s_{11}}{c \tanh(\rho_0)}\right)\nonumber\\ \times \left(1+\frac{\varepsilon_1}{\varepsilon_2} \tanh(3\rho_0)+\frac{4\pi\sigma i 3 s_{33}}{c k_0 \varepsilon_2 l \sinh(\rho_0)}  \right)\nonumber\\
- \left(\frac{4\pi\sigma}{c}\right)^2 \frac{\cosh(\rho_0) i s_{13} s_{31}}{k_0 \varepsilon_2 l \sinh^2(\rho_0)} 
\end{eqnarray}
Once the scattering amplitude $b_1$ is found, the polarizability $\alpha_x$ is given by (see  appendix \ref{dipolo})
%
%
\begin{equation}\label{alfax}
\alpha_x = \frac{i (a+b)}{4 k_0} \widetilde{b_1}. 
\end{equation}

Following the same steps that allow us to obtain Eq. (\ref{homogeneo5}) but taking the electric field along the $y$ axis, we obtain the lowest order dispersion equation for $y$ polarization, 
\begin{eqnarray}\label{homogeneo6}
 \left(1+\frac{\varepsilon_1}{\varepsilon_2}\coth(\rho_0)+\frac{4\pi\sigma}{c k_0 \varepsilon_2} i v_{11}\right) \left(1+\frac{\varepsilon_1}{\varepsilon_2}\coth(3\rho_0)+ 3 \frac{4\pi\sigma}{c k_0 \varepsilon_2} i v_{33}\right) \nonumber\\
+ 3 \left( \frac{4\pi \sigma}{c k_0 \varepsilon_2}\right)^2 (v_{13})^2=0,
\end{eqnarray}
\begin{eqnarray}\label{v_1k2}
v_{11}=\frac{1}{a} \left[ 1+\frac{3}{8} e^2\right],\nonumber\\
v_{13}=v_{31}=\frac{1}{a} \frac{1}{8} e^2,\nonumber\\
v_{33}=\frac{1}{a} \left[ 1+\frac{1}{4} e^2\right].
\end{eqnarray}
At the same order in eccentricity as we have written Eq. (\ref{homogeneo6}), the dipolar coefficient $d_1$ for the scattering of a plane wave polarized along $y$ axis, 
\begin{equation}\label{d1}
d_1 =E_0 e^{\rho_0} \widetilde{d_1}, 
\end{equation}
where $d_1$ is the dipolar amplitude of the scattered field in medium 2 for $y$ polarization and 
\begin{eqnarray}
\widetilde{d_1} = \frac{A_y}{q_1 q_3-N_{13} N_{31}},
\end{eqnarray}
where
\begin{eqnarray}
A_y=\left(ik_0 b [\varepsilon_2-\varepsilon_1]+\frac{4 \pi \sigma v_{11}}{c \coth(\rho_0)}\right)\nonumber\\
\times \left(1+\frac{\varepsilon_1}{\varepsilon_2} \coth(3\rho_0)+\frac{4\pi\sigma i 3 v_{33}}{c k_0 \varepsilon_2 l \cosh(\rho_0)}  \right)\\
- \left(\frac{4\pi\sigma}{c}\right)^2 \frac{\sinh(\rho_0) i v_{13} v_{31}}{k_0 \varepsilon_2 l \cosh^2(\rho_0)},
\end{eqnarray}
and
\begin{eqnarray}
q_k=1+\frac{\varepsilon_1}{\varepsilon_2} \coth(k\rho_0)+N_{kk},
\end{eqnarray}
\begin{eqnarray}
N_{mk}=\frac{4\pi\sigma}{c k_0 \varepsilon_2} i m v_{km}.
\end{eqnarray}
%
%
%
The corresponding polarizability $\alpha_y$ is given by (see appendix \ref{dipolo})
\begin{equation}\label{alfay}
\alpha_y = \frac{i (a+b)}{4 k_0} \widetilde{d_1}. 
\end{equation}

It is interesting to note that in the limit of $\rho_0 \rightarrow \infty$, the polarizabilities (\ref{alfax}) and (\ref{alfay}) tend to the value corresponding to the circular case,  
\begin{equation}\label{alfa_circulo}
\alpha_{c} = \varepsilon_2 \frac{b^2}{2} \left[  \frac{\varepsilon_1-\varepsilon_2+\frac{4\pi\sigma i}{ck_0 b}}{\varepsilon_1+\varepsilon_2+\frac{4\pi\sigma i}{ck_0 b}}\right].
\end{equation}
%

Polarizabilities (\ref{alfax}) and (\ref{alfay}) constitutes the central object for optical interactions where the  quasistatic regime is applicable \cite{mortensen}, such as enhanced and confined optical near--fields \cite{bonod}, metasurface and metagrating applications \cite{NV} and nanoparticles bonding \cite{NV1,NV2,kostina}. In the present work, we use these expressions to calculate the power scattered by the particle in the quasistatic limit as follows.  
Taking into account that the radiated power by a point  dipole $\mathbf{p}$ is \cite{cuevas5},
\begin{equation}\label{Potencia}
P =\frac{\pi \omega^3}{4 c^2} |\mathbf{p}|^2,
\end{equation}
and neglecting third order contributions,  we can replace the induced dipole moment components  $p_j=\alpha_j\,E_0$ ($j=x,\,y$) 
to calculate the  power scattered  by the elliptical particle in the dipolar limit. Considering that the incident power is $P_0=\frac{c}{8\pi} E_0^2 2 L_i$, $L_i=a,\,b$ for $x$ and $y$ polarization, respectively, the normalized power scattered by the particles is written as
\begin{equation}\label{Potenciaxy}
\frac{P_i}{P_0}=\frac{ \pi^2 \omega^3}{L_i c^3} |\alpha_i|^2.
\end{equation}
%


\section{Results}
In this section we use the formalism developed in the above section to calculate the eigenfrequencies and the scattering cross section curves. In all the examples the wire is immersed in vacuum ($\varepsilon_2=\mu_2=1$), the dielectric core has a permittivity $\varepsilon_1 = 3.9$ and permeability $\mu_1=1$. The graphene parameters are $T = 300$K and $\gamma_g = 0.1$ meV. 

In order to explore the effects that the departure from the circular geometry has on the dispersive characteristics of LSPs, we compare the results obtained for all elliptical shapes with those obtained in the circular case with the same perimeter. We assume that the perimeter is sufficiently large to describe the optical properties of the wires as characterized by the same local surface conductivity as planar graphene (see appendix \ref{grafeno}). Moreover, since the  aroused interest on graphene due to plasmonic  properties in THz and IR frequency regions,  graphene--coated wires with a micro-sized cross section have become an attractive platform for optical applications (see \cite{depineguias,DaTeng,nosich} and Refs. therein).  
In this way, we have chosen a perimeter  equal to $\pi\,\mu$m for all examples, excepting those presented in Fig. \ref{figura5}a where we have selected  $\pi/2\,\mbox{and}\,\pi/4\,\mu$m. Even though nonlocal effects could appear in our  system for frequencies lower than characteristic resonance frequencies, this is not interesting for our purposes (see appendix \ref{nolocal}).  


Firstly, we evaluate the coupling mechanism  between multipolar orders provided by the graphene current. To do this, by solving Eq. (\ref{nohomogeneo}) we calculate the amplitude modulus  $|b_m|$  for $m=1$ and $3$ (both non-null lowest order)  as a function of $\omega/c$ frequency and for three values of eccentricity, $a/b=1.05,\,1.2$ and $1.4$.  Without loss of generality we made the calculation for horizontal (or $x$) polarization (electric field parallel to $x$ axis). 
For the lowest eccentricity  value of $a/b=1.05$ we observe that the amplitude $b_1$ reaches it maximum value ($\approx 10$) at 
$\hbar\omega=31.52$meV corresponding to the dipolar excitation. At the same frequency, the amplitude $b_3$ reaches a local maximum ($\approx 0.1$), evidencing the  coupling between third and first order modes. This means that the amplitude $b_3$ not only contributes to the scattered field  at the resonance frequency of the $ m = 1$  order (where the $|b_1|$ amplitude reaches the absolute maximum value), a fact that also occur on metallic nanoparticles  \cite{Bohren,mertens,marocico,barthes} for which the field amplitudes are decoupled, but also reaches a local maximum at this frequency.

\begin{figure} [htbp!]
\centering
\resizebox{0.7\textwidth}{!}
{\includegraphics{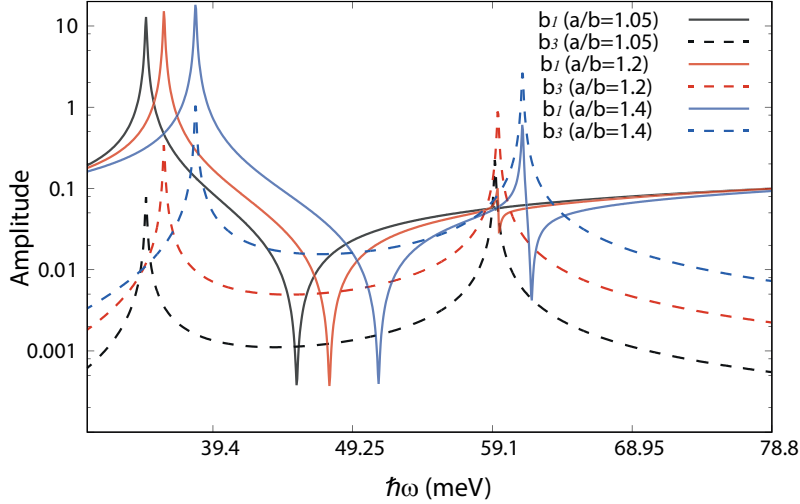}}
\caption{\label{fig:epsart2} Amplitudes $b_m$ modulus ($m=1$ and $3$)  as a function of $\omega/c$ frequency for $a/b=1.05,\,1.2,\,1.4$ and for $x$ polarization. The perimeter of the ellipse is $\pi\,\mu$m and chemical potential $\mu_g=0.5eV$. 
}\label{figura1}
\end{figure}

On the other hand, the amplitude $b_3$ reaches the absolute maximum value at 
$\hbar\omega \approx 59$meV corresponding to the hexapolar order excitation, while the first order amplitude $b_1$ shows a local maximum and minimum at   this frequency value. As eccentricity values are increasing, two effects are observed: on the one hand,   dipolar and hexapolar order frequencies moves to higher values and, on the other hand,  the coupling mechanism between different orders  appear more visible, as can be seen in Figure \ref{figura1} where the local maximum value of $b_3$ curve at the resonant  dipolar frequency  (where the $b_1$ amplitude reaches the absolute maximum value)  increases with the value of $a/b$. The same behavior is observed near $\hbar \omega \approx 59$meV, where the local variation of the $b_1$ amplitude at the hexapolar resonant frequency  is more visible with the $a/b$ increment. These results show how a non--zero surface conductivity on graphene  couples different orders of the same parity (see second term in Eq. (8))  causing all these modes to reach a local maximum at the resonant frequency of one of them.
 
In the vertical (or $y$) polarization (electric field parallel to $y$ axis) case (not shown in Figure \ref{figura1}), we obtain similar  results with the only exception that the resonant frequency is a decreasing function of the eccentricity $a/b$, instead of being increasing as occur in the $x$ polarization case.  

\begin{figure}
\centering
\resizebox{0.7\textwidth}{!}
{\includegraphics{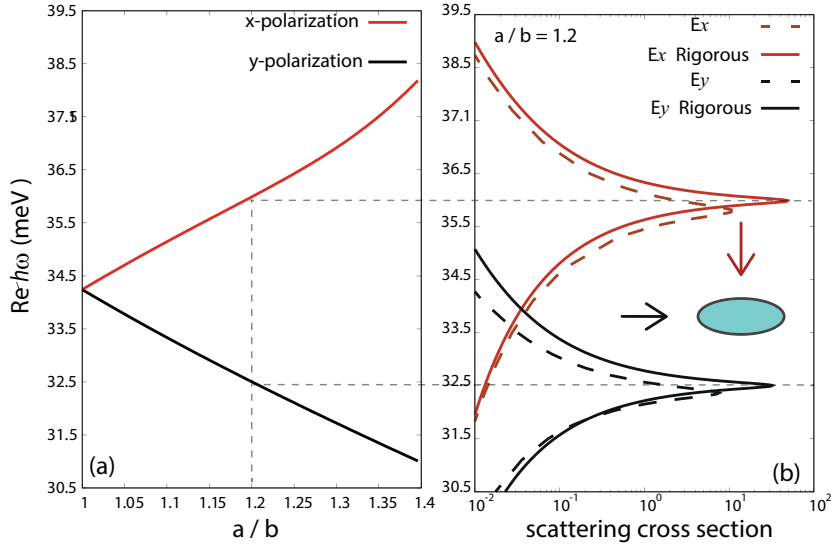}}
\caption{\label{fig:epsart3} (a) Real part of the eigenfrequencies as a function of $\frac{a}{b}$ ($\frac{a}{b}=\sqrt{e^2+1}$). (b) Scattering cross section for incidence direction parallel to $y$ axis (electric field along $x$ axis) and parallel to y axis (electric field along the $y$ axis).  Continuous lines correspond to the calculus using Eq. (\ref{Potenciaxy})  and dashed line curves correspond to calculation using GSIM. The arrows indicate the incident direction. The perimeter of the ellipse is $\pi\,\mu$m and chemical potential $\mu_g=0.5$eV. 
}\label{figura2}
\end{figure}

In order to gain insight about the resonant  frequency dependence with the eccentricity of the wire,  we solve  the homogeneous problem at lowest order in eccentricity to obtain the dispersion equation for lower modes. 
Figure \ref{figura2} shows the dispersion relation as a function of $a/b$ ($a/b \leq 1.4$) and the scattering curves for both horizontal 
and vertical 
polarization for $a/b=1.2$ ($e=0.66$). From Figure \ref{figura2}a we see two branches, the upper branch, calculated with Eq. (\ref{homogeneo5}), corresponds to $x$ polarization and the lower branch, calculated with Eq. (\ref{homogeneo6}), corresponds to $y$ polarization. For null eccentricity, $a=b$, these two branches converge to the value $\hbar \omega=34.28$meV corresponding to the case of circular cross section. As the value of eccentricity increases, the gap between two branches increases leading to an increment in the  splitting between resonance frequencies  when the structure is illuminated with a plane wave. In Figure \ref{figura2}b we plotted the scattering curves for plane wave incidence by using Eq. (\ref{Potenciaxy}) with $i=x$ and $i=y$ for $x$ and $y$  polarization, respectively, (continuous line). The calculation also was made by using a rigorous method based on the Green surface integral method (GSIM) (dashed line) \cite{valencia}. Two peaks are observed. The lower frequency peak, angle of  incidence $\theta=90^\circ$, corresponds to excitation of $y$-polarized LSPs, \textit{i.e.}, surface plasmons whose polarizability $\alpha_y$ is along the $y$ axis. The upper frequency peak, angle of  incidence  $\theta=0$, correspond to excitation of $x$-polarized LSPs,   \textit{i.e.}, surface plasmons whose polarizability $\alpha_x$ is along the $x$ axis. Both resonance frequencies agree well with  the values calculated from the dispersion equation. We also observe a small red shifting of the resonance peaks calculated using GSIM with respect to those calculated using Eq. (\ref{Potenciaxy}).  

It is worth noting that unlike metallic structures, where  the lower modal  frequency is associated to oscillations along the major axis (the $x$ axis in the present case) and the upper modal frequency is along the minor axis (the $y$ axis in the present case), from Figure \ref{figura2} we see that the situation is completely other way around for graphene wrapped wires. This fact is due to the geometrical difference between the charge oscillations on each of the systems. While in the metallic case the displacement of charge is along the induced electric field, in the graphene  particles case, the movement of charge is along the boundary elliptical  curvature. 

\begin{figure}
\centering
\resizebox{0.7\textwidth}{!}
{\includegraphics{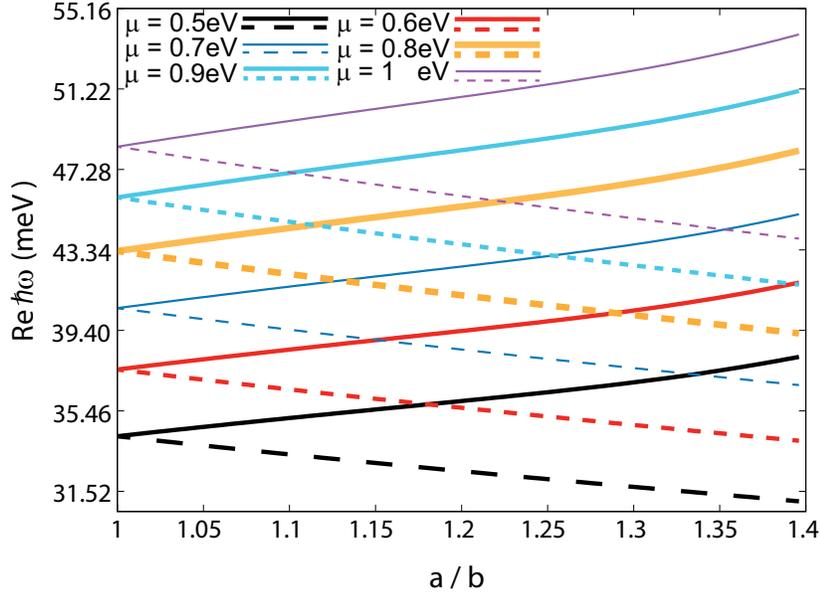}}
\caption{\label{fig:epsart3} Real part of the eigenfrequencies as a function of $\frac{a}{b}$ ($\frac{a}{b}=\sqrt{e^2+1}$) for $\mu_g=0.6,\,0.7,\,0.8,\,0.9,\,1$eV. Solid lines correspond to $x$ polarization and dashed lines corresponds to $y$ polarization.   Other parameters are the same as in Figure \ref{figura2}  
}\label{figura3}
\end{figure}

\begin{figure}
\centering
\resizebox{0.7\textwidth}{!}
{\includegraphics{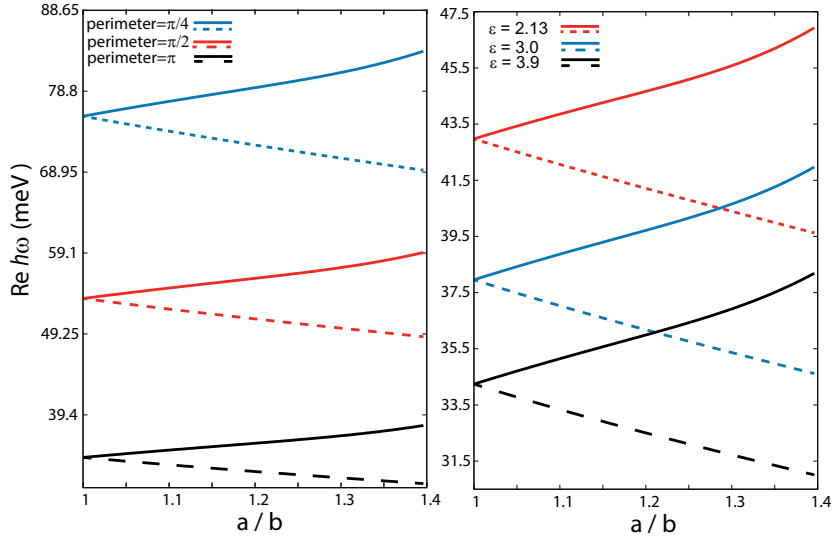}}
\caption{\label{fig:epsart3} Real part of the eigenfrequencies as a function of $\frac{a}{b}$ ($\frac{a}{b}=\sqrt{e^2+1}$).  (a) $\varepsilon_1=3.9$ and  various values of perimeters:   $\pi,\,\pi/2,\,\pi/4\,\mu$m, (b) for perimeter equal to $\pi\mu$m and various values of permittivities,   $\varepsilon_1=3.9,\,3,\,2.13$.  The chemical potential is $\mu_g=0.5$eV. Solid lines correspond to $x$ polarization and dashed lines corresponds to $y$ polarization.  
}\label{figura5}
\end{figure}

In order to study the behavior with  graphene parameters, we calculate the eigenfrequency dependence on chemical potential. Figure \ref{figura3} shows the eigenfrequency branches, one for $x$ polarization and the other for $y$ polarization, for various values of the  chemical potential, $\mu_g=0.6,\,0.7,\,0.8,\,0.9,\,1$eV. We observe that eigenfrequencies of the upper and lower branches increase with the chemical potential increment. This is consistent with the fact that frequency plasmonic resonances for circular cross section cylinders are proportional to $\sqrt{\mu_g}$ \cite{CRD}.

Finally, we calculate the dependence of the eigenfrequencies with the size and the  contrast between external and
internal constitutive parameters. In Figure \ref{figura5}a we observe that both  branches,  corresponding to $x$ and $y$ polarizations,  increase their frequencies as the ellipse perimeter is decreased from $\pi$ to $\pi/4\,\mu$m. This fact can be understood from the circular case studied in \cite{CRD}, where  we have demonstrated that the eigenfrequency depends on the radius as $\omega\approx R^{-1/2}$. In Figure \ref{figura5}b we have plotted the eigenfrequency branches for three values of the permittivity of the internal medium,  $\varepsilon_1=3.9,\,3,\,2.13$. We observe that Re$\,\hbar\omega$ increases with the decreasing of the permittivity value. This behavior is consistent with the fact that  the eigenfrequency depends on permittivities as $\omega \approx (1+\varepsilon_1)^{-1/2}$ for circular graphene wires \cite{CRD}.

\section{Conclusion}

In conclusion, we have analytically studied  the dispersive characteristics  for an ellipical dielectric wire wrapped with graphene.  We have  found two branches corresponding to the lowest frequency band, one of them corresponds to eigenmodes with a dipole moment along the major axis  and the other with  a dipole moment along the minor axis of the wire elliptical cross section. Interestingly, we found that contrary to what happens in the metallic plasmonic case, the low dipolar frequency branch corresponds to polarizability oscillations along minor ellipse axis and the high dipolar frequency eigenmode corresponds to polarizability oscillations along the major ellipse axis. We have found analytical expressions for the polarizability elements along the ellipse axis which reduces to that of the circular shaped wire as the eccentricity parameter tends to zero.   

We think that the results provided in this work can be employed  for a deeper interpretation of the dispersive characteristics of elliptical graphene plasmons,  which opens up possibilities for practical applications using  structures capable to manipulate  light at sub-wavelength scale. 

\appendix

\section{Graphene conductivity} \label{grafeno}
\setcounter{equation}{0}
\renewcommand{\theequation}{A{\arabic{equation}}}

The  graphene layer is considered as an infinitesimally thin layer with a frequency-dependent surface conductivity $\sigma(\omega)$ given by the Kubo formula \cite{falko}, $\sigma= \sigma^{intra}+\sigma^{inter}$, with the intraband and interband contributions given by
\begin{equation} \label{intra}
\sigma^{intra}(\omega)= \frac{2i e^2 k_B T}{\pi \hbar^2 (\omega+i\gamma_g)} \mbox{ln}\left[2 \mbox{cosh}(\mu_g/2 k_B T)\right],
\end{equation}  
\begin{eqnarray} \label{inter}
\sigma^{inter}(\omega)= \frac{e^2}{\hbar} \Bigg\{   \frac{1}{2}+\frac{1}{\pi}\mbox{arctan}\left[(\hbar \omega-2\mu_g)/2k_BT\right]-\nonumber\\
\frac{i}{2\pi}\mbox{ln}\left[\frac{(\hbar \omega+2\mu_g)^2}{(\hbar \omega-2\mu_g)^2+(2k_BT)^2}\right] \Bigg\},
\end{eqnarray}  
where $\mu_g$ is the chemical potential (controlled with the help of a gate voltage), $\gamma_g$ the carriers scattering rate, $e$ the electron charge, $k_B$ the Boltzmann constant and $\hbar$ the reduced Planck constant.

\section{Boundary conditions} \label{desarrollo1}
\setcounter{equation}{0}
\renewcommand{\theequation}{B{\arabic{equation}}}

By replacing Eqs. (\ref{campo_dentro}) and (\ref{campo_fuera}) into Eq. (\ref{CC}) we obtain a set of two coupled equations for amplitudes $a_m$ and $b_m$,
\begin{equation}\label{CC1}
\sum_{m=1}^{+\infty} \left( \frac{a_m}{\varepsilon_1} \cosh(m\rho_0) m + m\frac{b_m}{\varepsilon_2} e^{-m\rho_0}\right) \sin(\phi)=\frac{1}{\varepsilon_2} \frac{\partial H^{(inc)}}{\partial \rho}
\end{equation}
\begin{eqnarray}\label{CC2}
\sum_{m=1}^{+\infty} \left( b_m e^{-m\rho_0}-a_m \sinh(m\rho_0) + \frac{4\pi \sigma }{c k_0 \varepsilon_2}i m b_m e^{-m\rho_0} \frac{1}{f(\rho_0,\phi)}  \right) \nonumber \\
\times \sin(m\phi)   =\nonumber \\
-H^{(inc)}|_{\rho_0}+ \frac{4\pi \sigma }{c k_0 \varepsilon_2}i \frac{1}{ f(\rho_0,\phi)} \frac{\partial H^{(inc)}}{\partial \rho}|_{\rho_0}.
\end{eqnarray}
Note that the left hand side of Eq. (\ref{CC1}) is written as a sine Fourier series, then we can expand the right hand side as a sine series and find, performing Fourier integral, a lineal equation between amplitudes $a_m$ and $b_m$,
\begin{equation}\label{CC1c}
  \frac{a_k}{\varepsilon_1} \cosh(k\rho_0) k + \frac{b_k}{\varepsilon_2} k e^{-k\rho_0} =-i k_0 E_0 l \cosh(\rho_0) \delta_{k 1}.
\end{equation}
Contrary,  Eq. (\ref{CC2}) presents a difficulty related with the graphene  current  factor $f^{-1}(\rho_0,\phi)$.  Multiplying Eq. (\ref{campo_fuera}) by $\sin(k \phi)$ ($k \geq 1$) and integrating in $[-\pi,\,\pi]$, we obtain,
\begin{eqnarray}\label{CC2a}
 b_k e^{-k\rho_0}-a_k \sinh(k \rho_0) + \sum_{m=1}^{+\infty}  \frac{4\pi \sigma }{c k_0 \varepsilon_2}i m b_m e^{-m\rho_0} s_{km}   =\nonumber \\
-h_k+ \frac{4\pi \sigma }{c k_0 \varepsilon_2} i dh_k, 
\end{eqnarray}
where
\begin{eqnarray}\label{s_mk}
s_{km}=\frac{1}{\pi}\int_{-\pi}^{\pi} \frac{1}{ f(\rho_0,\phi)} \sin(m\phi) \sin(k \phi) d\phi 
\end{eqnarray}
\begin{eqnarray}\label{hk}
h_{k}=\frac{1}{\pi}\int_{-\pi}^{\pi}  H^{(inc)}|_{\rho} \sin(k \phi) d\phi 
\end{eqnarray}
\begin{eqnarray}\label{dhk}
dh_k=\frac{1}{\pi}\int_{-\pi}^{\pi} \frac{1}{ f(\rho_0,\phi)} \frac{\partial H^{(inc)}}{\partial \rho}|_{\rho_0} \sin(k \phi) d\phi. 
\end{eqnarray}
Note that both functions $H^{(inc)}|_{\rho_0} $ and $\frac{\partial H^{(inc)}}{\partial \rho}|_{\rho_0}$ have an angular dependence $\approx \sin(\phi)$ and thus $h_k \approx \delta_{k 1}$ and $dh_k \approx s_{km} \delta_{m1}$. Therefore, Eq. (\ref{CC2a}) can be    written as,
\begin{eqnarray}\label{CC2b}
 b_k e^{-k\rho_0}-a_k \sinh(k \rho_0) + \sum_{m=1}^{+\infty}  \frac{4\pi \sigma }{c k_0 \varepsilon_2} i m b_m e^{-m\rho_0} s_{km}   =\nonumber \\
i k_0 E_0 \varepsilon_2 l \sinh(\rho_0) \delta_{k 1}+ \frac{4\pi \sigma }{c} E_0  \coth(\rho_0) s_{k1}. 
\end{eqnarray}
%
Note that in absence of graphene, \textit{i.e.} $\sigma=0$, the system of Eqs. (\ref{CC1c}) and (\ref{CC2b}) are uncoupled for field amplitudes. Contrary, in presence of graphene, the graphene current term in the left hand side of Eq. (\ref{CC2b}) couples  orders with the same parity. This is true because the matrix elements $s_{km}\not=0$ provided that $m+k$ be  even (Eq. (\ref{s_mk})).

By eliminatig the amplitudes $a_m$  from Eqs. (\ref{CC1c}) and (\ref{CC2b}), we 
obtain  an equation for amplitudes $b_m$, 
\begin{eqnarray}\label{CC2d}
 b_k e^{-k\rho_0} \left(1+\frac{\varepsilon_1}{\varepsilon_2} \tanh(k\rho_0)+ \frac{4\pi \sigma }{c k_0 \varepsilon_2} i k s_{kk} b_k e^{-k\rho_0}\right)+\nonumber\\ 
 \sum_{m\not=k}^{+\infty}  \frac{4\pi \sigma }{c k_0 \varepsilon_2} i m b_m e^{-m\rho_0} s_{km}   =\nonumber \\
i k_0 E_0 (\varepsilon_2-\varepsilon_1) l \sinh(\rho_0) \delta_{k 1}+ \frac{4\pi \sigma }{c} E_0  \coth(\rho_0) s_{k1}. 
\end{eqnarray}

\section{Explicit form of $s_{k m}$ matrix elements} \label{orden}
\setcounter{equation}{0}
\renewcommand{\theequation}{C{\arabic{equation}}}

We expand  the current  factor as power of $\sin(\phi)$,
\begin{eqnarray}\label{desarrollo}
\frac{1}{h(\rho_0,\phi)}=\frac{1}{\sinh(\rho_0) \sqrt{1+\sin^2(\phi) e^2}}=\nonumber\\ \frac{1}{\sinh(\rho_0)} [1-\sin^2(\phi)\frac{1}{2} e^2+\frac{3}{8}\sin^4(\phi) e^4 -\frac{5}{16}\sin^6(\phi) e^6  \nonumber\\ 
+\frac{35}{128}\sin^8(\phi) e^8
-\frac{63}{226}\sin^{10}(\phi) e^{10}\nonumber\\
+\frac{231}{1024}\sin^{12}(\phi) e^{12}-\frac{429}{2048}\sin^{14}(\phi) e^{14}+...].
\end{eqnarray}
%
Note that the expansion (\ref{desarrollo}) is valid for $a/b<\sqrt{2}$. By replacing Eq. (\ref{desarrollo}) into Eq. (\ref{s_mk}), we obtain
\begin{eqnarray}\label{s_mk2}
s_{km}=\frac{1}{b \pi}\int_{-\pi}^{\pi} [1-\sin^2(\phi)\frac{ 1}{2} e^2+\frac{3}{8}\sin^4(\phi) e^4\nonumber\\
-\frac{5}{16}\sin^6(\phi) e^6+\frac{35}{128}\sin^8(\phi) e^8 
-\frac{63}{226}\sin^{10}(\phi) e^{10}\nonumber\\
+\frac{231}{1024}\sin^{12}(\phi) e^{12}-\frac{429}{2048}\sin^{14}(\phi) e^{14}+...] \nonumber\\
\times \sin(m\phi) \sin(k \phi) d\phi. 
\end{eqnarray}
From this equation, we can see that $s_{km} \approx e^2$ ($s_{km}$ is of order $O(2)$) for $k \not= m$. For example, if we    
 consider $k=1$, 
 it is straightforward verify,
\begin{eqnarray}\label{s_1k}
s_{1m}=\frac{1}{b} [\delta_{m1}-\frac{e^2}{2}\left(\frac{3}{4}\delta_{m1}-\frac{1}{4}\delta_{m3}\right) \nonumber\\
+\frac{3}{8} e^4 \left(\frac{5}{8}\delta_{m1}-\frac{5}{16}\delta_{m3}+\frac{11}{16}\delta_{m5}\right) + ...].
\end{eqnarray}
We can see that $s_{11}$ is of order $O(0)$, $s_{13}$ is of order $O(2)$, $s_{15}$ is order $O(4)$, ..., $s_{1m}$ is order $O(m-1)$. 
%
%

\section{Field of a dipole moment placed at the origin} \label{dipolo}
\setcounter{equation}{0}
\renewcommand{\theequation}{D{\arabic{equation}}}

We consider a line dipole source (whose axis lies along the $z$ axis)  with a dipole moment $\mathbf{p}$ is placed at the origin. The magnetic field is along the $z$ axis ($\mathbf{H} (\mathbf{r}) = \phi(\mathbf{r}) \hat{r}$). The wave equation for $\phi( \mathbf{r} )$ when the retardation is
negligible reads 
\begin{eqnarray}\label{laplace}
\nabla^2 \phi(\mathbf{r}) = -4\pi i k_0 \mathbf{p} \times \nabla \delta(\mathbf{r}).
\end{eqnarray}
The solution $\phi(\mathbf{r})$ of Eq. (\ref{laplace}) can be written as
\begin{eqnarray}\label{laplace2}
\phi(\mathbf{r}) =  i k_0 \mathbf{p} \times \nabla f(\mathbf{r}),
\end{eqnarray}
where $f(\mathbf{r})$ satisfies 
\begin{eqnarray}\label{laplace3}
\nabla^2 \phi(\mathbf{r}) = -4\pi \delta(\mathbf{r}).
\end{eqnarray}
By solving Eq. (\ref{laplace3}) we obtain $\phi(\mathbf{r})=-2 \log(\mathbf{r})$. Then, the magnetic field $\phi(\mathbf{r})$ of a point dipole calculated using Eq. (\ref{laplace2}) in elliptical coordinates (\ref{coordenadas}) is given by
\begin{eqnarray}\label{laplace4}
\phi(\mathbf{r}) = -2 i k_0 \frac{p_x \sinh{\rho} \sin\phi-p_y \cosh{\rho} \cos\phi}{l(\cosh^2{\rho}-\sin^2{\phi} )},
\end{eqnarray}
where $p_x$ and $p_y$ are the Cartesian components of the dipole moment $\mathbf{p}$. For $\rho$ values large enough, this expression can be rewritten as  
\begin{eqnarray}\label{laplace5}
\phi(\mathbf{r}) = -4 i k_0 \frac{p_x  \sin\phi-p_y  \cos\phi}{l} e^{-\rho}.
\end{eqnarray}
By comparing this equation (with $p_y=0$) with the first term in the scattered field given in  Eq. (\ref{campo_fuera}) 
we obtain 
the induced electrical dipole as a function of the $b_1$ scattering amplitude 
\begin{equation}\label{px}
p_x =\frac{i l}{4 k_0}b_1= \frac{i l}{4 k_0} E_0 e^{\rho_0} \widetilde{b_1}=\frac{i (a+b)}{4 k_0} E_0  \widetilde{b_1}, 
\end{equation}
where the sub-index stand for the direction of the dipole moment. In the last equality in Eq. (\ref{px})  we have used $l e^{\rho_0}=a+b$. 
%
Finally, dividing Eq. (\ref{px}) by $E_0$, the corresponding polarizability $\alpha_x$ is obtained. 

In the similar way, by comparing Eq. (\ref{laplace5}) (with $p_x=0$) with the second term in Eq. (\ref{campo_fuera}) we obtain the  induced electrical dipole  
\begin{equation}\label{py}
p_y =\frac{i l}{4 k_0}d_1= \frac{i l}{4 k_0} E_0 e^{\rho_0} \widetilde{d_1}=\frac{i (a+b)}{4 k_0} E_0  \widetilde{d_1}. 
\end{equation}
By dividing Eq. (\ref{py}) by the incident field amplitude $E_0$, we obtain the $\alpha_y$ polarizability (\ref{alfay}).

\section{Nonlocal response in graphene conductivity} \label{nolocal}
\setcounter{equation}{0}
\renewcommand{\theequation}{E{\arabic{equation}}}

\begin{figure} [htbp!]
\centering
\resizebox{0.7\textwidth}{!}
{\includegraphics{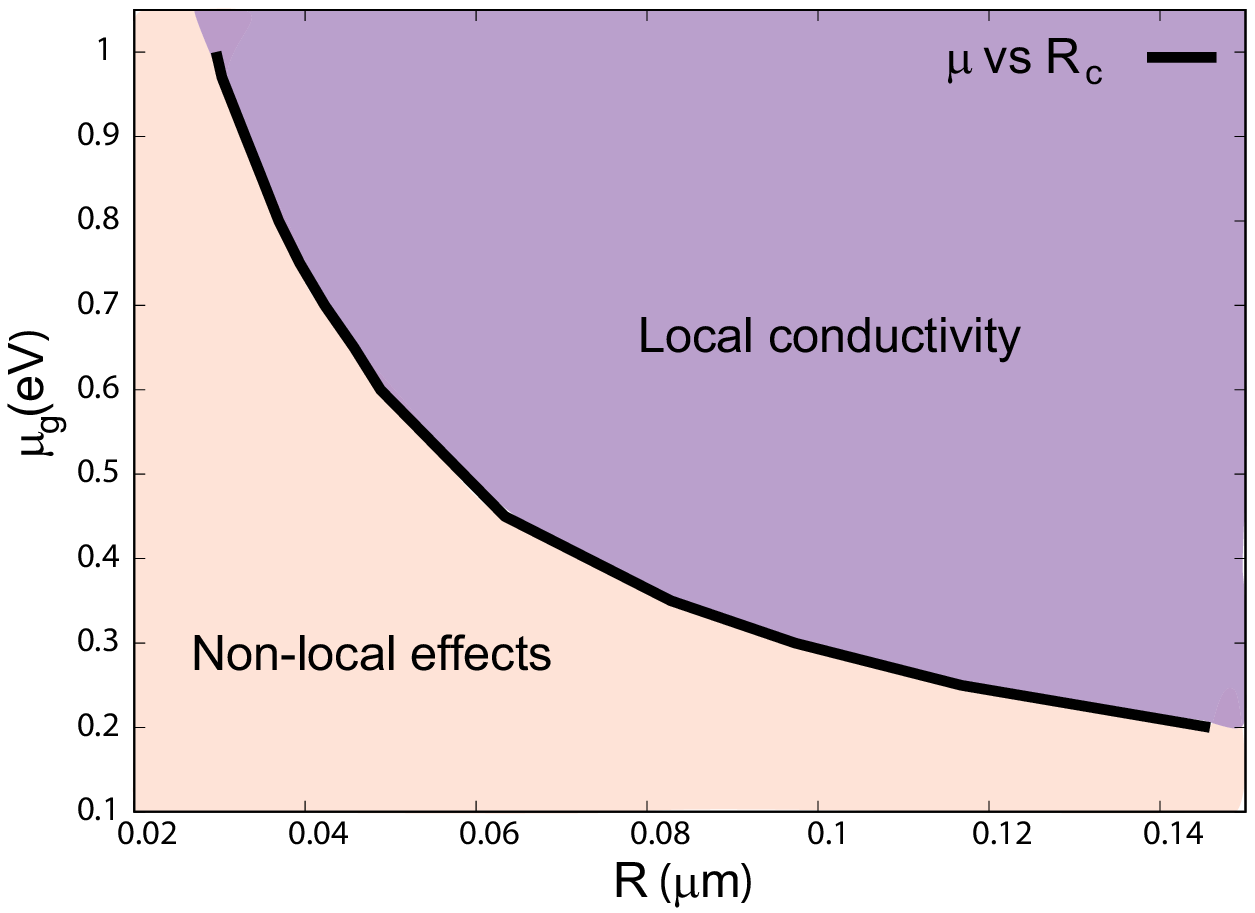}}
\caption{\label{fig:epsart2}  Regions of the plane $\mbox{R}-\mu_g$ (radius and chemical potential)  where, according to Eq. (\ref{omegaR}), local and non local effects take  place. The curve separating these two regions corresponds to critical values of the radius, $R_c$, that verify  the equality in Eq. (\ref{omegaR}) for varying $\mu_g$ values.
}\label{mu_vs_R}
\end{figure}

The objective of this section is to provide information about the relation between the local graphene conductivity response  considered in this work and the dimensions of the proposed structure. 
The local approximation begins to break down as a reduction of the structure takes place.   
 The nonlocality or spatial dispersion in graphene conductivity 
arises when the plasmon phase velocity $v_{SP}$ is slow and  comparable with the electron Fermi velocity $v_{F} \approx 10^{12}\, \mu m/s$. Since, we focused on cylindrical  structures whose cross section is a slight deviation from the circular cross section, an estimation of their size can be made by considering a cylinder with circular cross section and radius $R$ ($a<R<b$). 
Taking into account the small size of the cylinder, $R/\lambda<<1$ ( $\lambda=2 \pi / \omega$ the wavelength), the phase velocity of plasmons $v_{SP}=k_{SP}/\omega$ can be estimated by using the quasistatic approximation. In this way, the surface plasmon  effective momentum  $k_{SP}$, which is  along the azimutal angle  ($\phi$ axis), can be written as \cite{CRD},   
\begin{eqnarray}\label{k_sp}
k_{SP}=\frac{m}{R},
\end{eqnarray}
where $m$ is the eigenmode order. If we consider $m=1$ (dipolar order), the factor
\begin{eqnarray}\label{v_F_v}
\frac{v_F}{v_{SP}}=\frac{v_F}{R \, \omega} \approx \frac{10^{12} \mu \mbox {m}/s}{R \, \frac{\omega}{c} \, 3 \times 10^{14}  \mu m/s } =  \frac{1}{R \, \frac{\omega}{c} \, 300 }. 
\end{eqnarray}
%
The smaller $\frac{v_F}{v_{SP}}$ in  Eq.  (\ref{v_F_v}), the better the local conductivity approximation. %
For example, if we consider a value $\frac{v_F}{v_{SP}}<1/10$, graphene  local conductivity differs less than 1$\%$ from that considering non local effects \cite{mortensen}. As a consequence, our method is acceptable provided that
\begin{eqnarray}\label{omegaR}
R \, \frac{\omega}{c}>  \frac{1}{30}. 
\end{eqnarray}
Equation (\ref{omegaR}) establishes a reasonable limit for the applicability of our model. Figure  \ref{mu_vs_R} shows a map with two regions. In one of them the nonlocal effects can be neglected (upper region), whereas in the other (lower region) these effects become important. The curve separating these regions corresponds to $R$ values, named $R_c$, for which the equality in  Eq. (\ref{omegaR}) is fulfilled.

\section*{Acknowledgment}
The authors acknowledge the financial supports of Universidad Austral O04-INV00020  and 
Consejo Nacional de Investigaciones Cient\'{\i}ficas y T\'ecnicas (CONICET).

\end{document}